\documentclass[11pt]{article}

\usepackage{amsfonts,amsthm,amsmath,amssymb,bbm,mathrsfs,verbatim,paralist,mathtools}
\usepackage{epsfig,color}
\usepackage{xcolor}
\usepackage[sort&compress,numbers]{natbib}
\usepackage{graphicx}
\usepackage[font=small,labelfont=bf]{caption}

\numberwithin{equation}{section}

\newcommand{\be}{\begin{equation}}
\newcommand{\ee}{\end{equation}}
\newcommand{\bea}{\begin{align}}
\newcommand{\eea}{\end{align}}

\newcommand{\abs}[1]{\lvert#1\rvert}

\def\rw{\rightarrow}

\def\a{\alpha}

\makeatletter
\def\underbracket{\@ifnextchar [
  {\@underbracket} {\@underbracket [\@bracketheight]}}
\def\@underbracket[#1]{\@ifnextchar [
  {\@under@bracket[#1]} {\@under@bracket[#1][0.2em]}}
\def\@under@bracket[#1][#2]#3{
  \mathop {\vtop {\m@th \ialign {##
           \crcr $\hfil \displaystyle {#3}\hfil$
           \crcr \noalign {\kern 3\p@ \nointerlineskip }\upbracketfill 
{#1}{#2}
           \crcr \noalign {\kern 3\p@ }
           \crcr }}}\limits}
\def\upbracketfill#1#2{$\m@th \setbox \z@ \hbox {$\braceld$}
                 \edef\@bracketheight{0.5pt}
                 \upbracketend{#1}{#2}
                 \leaders \vrule \@height #1 \@depth \z@ \hfill
                 \leaders \vrule \@height #1 \@depth \z@ \hfill
                 \upbracketend{#1}{#2}$}
\def\upbracketend#1#2{\vrule height #2 width #1\relax}

\def\overbracket{\@ifnextchar [
{\@overbracket} {\@overbracket [\@bracketheight]}}
\def\@overbracket[#1]{\@ifnextchar [
{\@over@bracket[#1]} {\@over@bracket[#1][0.2em]}}
\def\@over@bracket[#1][#2]#3{
\mathop {\vbox {\m@th \ialign {##
 \crcr \noalign {\kern 3\p@ }
 \downbracketfill {#1}{#2}
 \crcr \noalign {\kern 3\p@ \nointerlineskip }
 \crcr  $\hfil \displaystyle {#3}$
 \crcr
 } }}\limits}
\def\downbracketfill#1#2{$\m@th
 \setbox \z@ \vbox {$\braceld$}
 \edef\@bracketheight{0.5pt}
 \downbracketend{#1}{#2}
 \leaders \vrule \@height #1 \@depth \z@ \hfill
 \leaders \vrule \@height #1 \@depth \z@ \hfill
\downbracketend{#1}{#2}$}
\def\downbracketend#1#2{\vrule depth #2 width #1\relax}
\makeatother

\def\ce{competition effect}
\def\dn{distortion}

\newcommand{\Comp}{{\mathcal C}}
\newcommand{\Dist}{{\mathcal D}}

\title{Remarks on the energy costs of insulators in enzymatic cascades}
\author{{John Barton$^{1,2}$ and Eduardo D. Sontag$^3$}\\ \\
{\small $^1$ Departments of Chemical Engineering and Physics, MIT,}\\
{\small Cambridge, MA 02139 USA}\\ 
{\small $^2$ Ragon Institute of Massachusetts General Hospital,}\\
{\small MIT, and Harvard University, Boston, MA 02129 USA} \\
{\small $^3$ Department of Mathematics, Rutgers University,}\\
{\small Piscataway, NJ 08854-8019 USA}
}

\begin{document}

\maketitle




\section{Introduction} \label{introduction}

The connection between optimal biological function and energy use, measured for example by the rate of metabolite consumption, is a current topic of interest in the systems biology literature which has been explored in several different contexts. In \citep{biophysj}, we related the metabolic cost of enzymatic futile cycles with their capacity to act as \emph{insulators} which facilitate modular interconnections in biochemical networks. There we analyzed a simple model system in which a signal molecule regulates the transcription of one or more target proteins by interacting with their promoters. In this note, we consider the case of a protein with an active and an inactive form, and whose activation is controlled by the signal molecule. As in the original case, higher rates of energy consumption are required for better insulator performance.

\section{Mathematical model} \label{model}

Our focus is on biological pathways which transmit a single, time-dependent signal to one or more downstream targets. As a basic system we consider a signaling molecule Z which propagates a signal downstream by catalyzing the phosphorylation of a protein P, converting it to an active form P$^*$. The reactions describing this system are
\begin{align} \label{basic reactions}
\begin{aligned}
&\emptyset  \overset{k(t)}{\rw} \mbox{Z} \overset{\delta}{\rw} \emptyset\, , \\
&\mbox{Z+P} \overset{\a_1}{\underset{\a_2}{\rightleftharpoons}} \mbox{C}_1 \overset{k_1}{\rw} \mbox{Z+P}^*\, , \\
&\mbox{P}^* \overset{\delta}{\rw} \mbox{P}\, .
\end{aligned}
\end{align}
We refer to this system, in which the signaling molecule Z interacts directly with its target P, as the ``direct coupling'' (DC) system.
Here we have used a two-step model for the phosphorylation reaction, and set the small rate of the reverse reaction Z\,+\,P$^*\!\!\rw$\;C$_1$ to zero.
We assume that Z is produced or activated at rate
\be
k(t)=k\left(1+\sin{\omega}t\right)\nonumber\,,
\ee
and that its decay rate $\delta$ is constant in time. The total concentration of the protein P, $P_{\rm tot}$, is taken to be fixed. Here we assume that the phosphorylated protein P$^*$ decays to its dephosphorylated form P at the same rate as the decay rate of Z, but the results we present are not sensitive to this particular parameterization. The dynamics of the system \eqref{basic reactions} is then described by the ODEs
\begin{align} \label{basic ODEs}
\begin{aligned}
\frac{d Z}{d t}   &= k(t) - \delta Z - \a_1 Z \left(P_{\rm tot} - P^* - C_1\right) + \left(\a_2 + k_1\right) C_1 \, , \\
\frac{d C_1}{d t} &= \a_1 \left(P_{\rm tot} - P^* - C_1\right)Z - \left(\a_2 + k_1\right) C_1 \, , \\
\frac{d P^*}{d t} &= k_1 C_1 - \delta P^* \, .
\end{aligned}
\end{align}

In this description of the dynamics, the connection between the ``input,'' given by the time-dependent concentration $Z(t)$, and the ``output,'' which we take to be the concentration $P^*(t)$, is not perfectly modular. That is, due to the interaction between Z and P, given by the term $- \a_1 Z \left(P_{\rm tot} - P^* - C_1\right) + \left(\a_2 + k_1\right) C_1$ in \eqref{basic ODEs}, $Z(t)$ does not evolve in time as it would in isolation. Consequently, the output is also different than what would be expected if the communication from input to output was ``one-way.'' The term \emph{retroactivity} has been introduced to refer to these effects \citep{saez, DelVecchio:2008gy}, which are analogous to non-zero output impedance in electrical and mechanical systems.

\subsubsection*{Measures of retroactivity}

In \citep{biophysj} we introduced two new measures of retroactivity. The \emph{\dn} quantifies the difference between the actual output and what would be observed in an ``ideal'' system, free of retroactivity effects (represented in \eqref{basic ODEs} by the term $- \a_1 Z \left(P_{\rm tot} - P^* - C_1\right) + \left(\a_2 + k_1\right) C_1$ in the equation for $\frac{d Z}{d t}$), with dynamics given by
\begin{align} \label{ideal ODEs}
\begin{aligned}
\frac{d Z}{d t}   &= k(t) - \delta Z \, , \\
\frac{d C_1}{d t} &= \a_1 \left(P_{\rm tot} - P^* - C_1\right)Z - \left(\a_2 + k_1\right) C_1 \, , \\
\frac{d P^*}{d t} &= k_1 C_1 - \delta P^* \, .
\end{aligned}
\end{align}
The {\dn} defined as the time-averaged difference between the output in the real and ideal systems, normalized by the standard deviation of the ideal,
\begin{equation} \label{dist}
\Dist = \frac{1}{\sigma_{P^*_{\rm ideal}}} \, \langle \abs{P^*_{\rm ideal}(t) - P^*_{\rm real}(t)} \rangle \, .
\end{equation}
Here $\langle \cdot \rangle$ represents the time average, and $\sigma_{P^*_{\rm ideal}}$ the standard deviation
\begin{equation} \label{sigma Pideal}
\sigma_{P^*_{\rm ideal}} = \sqrt{\langle (P^*_{\rm ideal}(t)-\langle P^*_{\rm ideal}(t)\rangle)^2 \rangle} \, .
\end{equation}
Distortion is a measure of the faithfulness of signal transmission in the real system.

The second measure of retroactivity we considered is \emph{\ce}, which describes the change in the output in a fixed target as another target is added in parallel. We define this to be
\begin{equation} \label{comp}
\Comp = \frac{1}{\sigma_{P^*}}\,\left\langle \left| \left. \left(\frac{\partial \, P^*(t)}{\partial P^\prime_{\rm tot}}\right) \right|_{P^\prime_{\rm tot}=0} \right| \right\rangle \, ,
\end{equation} 
where P$^\prime$ is an additional target with total concentration $P_{\rm tot}^\prime$, which interacts with Z in the same way as P,
\begin{align}
\begin{aligned}
&\mbox{Z+P}^\prime \overset{\a^\prime_1}{\underset{\a^\prime_2}{\rightleftharpoons}} \mbox{C}^\prime_1 \overset{k^\prime}{\rw} \mbox{Z+P}^{\prime *}\, , \\
&\mbox{P}^{\prime *} \overset{\delta^\prime}{\rw} \mbox{P}^\prime\, .
\end{aligned}
\end{align}
As before, $\Comp$ is normalized by the standard deviation of the output
\begin{equation} \label{sigma P}
\sigma_{P^*} = \sqrt{\langle (P^*(t)-\langle P^*(t)\rangle)^2 \rangle} \, .
\end{equation}
The {\ce} quantifies the robustness of the output to changes in downstream targets.

\subsubsection*{Model of an insulator} \label{insulator}

As suggested in \citep{DelVecchio:2008gy} and explored in \citep{biophysj}, it is possible to reduce the effects of retroactivity by inserting an ``insulator'' between the input and output, implemented through a phosphorylation-dephosphorylation cycle. For the case considered here, the modified system including an insulator is described by the reactions
\begin{align} \label{PD reactions}
\begin{aligned}
& \emptyset \overset{k(t)}{\rw} \mbox{Z} \overset{\delta}{\rw} \emptyset \, , \\
& \mbox{Z+X} \overset{\beta_1}{\underset{\beta_2}{\rightleftharpoons}} \mbox{C}_2 \overset{k_2}{\rw} \mbox{X}^*\mbox{+Z} \, , \\
& \mbox{Y+X}^* \overset{\gamma_1}{\underset{\gamma_2}{\rightleftharpoons}} \mbox{C}_3 \overset{k_3}{\rw} \mbox{X+Y} \, , \\
& \mbox{X}^*\mbox{+P} \overset{\alpha_1}{\underset{\alpha_2}{\rightleftharpoons}} \mbox{C}_1 \overset{k_1}{\rw} \mbox{X}^*\mbox{+P}^* \, , \\
&\mbox{P}^* \overset{\delta}{\rw} \mbox{P}\, .
\end{aligned}
\end{align}
Here Z acts as a kinase, phosphorylating an intermediate signaling molecule X, whose active (phosphorylated) form in turn catalyzes the phosphorylation of P. Y is a phosphatase driving the dephosphorylation of X$^*$. We assume the total concentrations $X_{\rm tot}$ and $Y_{\rm tot}$ are fixed. The differential equations corresponding to the reactions in \eqref{PD reactions} are then
\begin{align} \label{PD ODEs}
\begin{aligned}
&\frac{d Z}{d t}    	= k(t)-\delta Z - \beta_1 Z \left(X_{\rm tot} - C_1 - C_2 - C_3\right) + \left(\beta_2 + k_2\right) C_2 , \\
&\frac{d C_2}{d t}	= \beta_1 Z \left(X_{\rm tot} - C_1 - C_2 - C_3\right) - \left(\beta_2 + k_2\right) C_2 , \\
&\frac{d C_3}{d t}	= \gamma_1 X^* \left(Y_{\rm tot} - C_3\right) - \left(\gamma_2 + k_3\right) C_3 , \\
&\frac{d X^*}{d t}	= k_2 C_2 - \gamma_1 X^* \left(Y_{\rm tot} - C_3\right) + \gamma_2 C_3 - \alpha_1 X^* \left(P_{\rm tot} - P^* - C_1\right) + \left(\alpha_2 +k_1\right) C_1 , \\
&\frac{d C_1}{d t}	= \alpha_1 X^* \left(P_{\rm tot} - P^* - C_1\right) - \left(\alpha_2 + k_1\right) \, C_1 \, , \\
&\frac{d P^*}{d t}	= k_1 C_1 - \delta \, P^* \, .
\end{aligned}
\end{align} 

\begin{figure}
\begin{center}
\hspace*{-40pt}
\includegraphics[width=16cm]{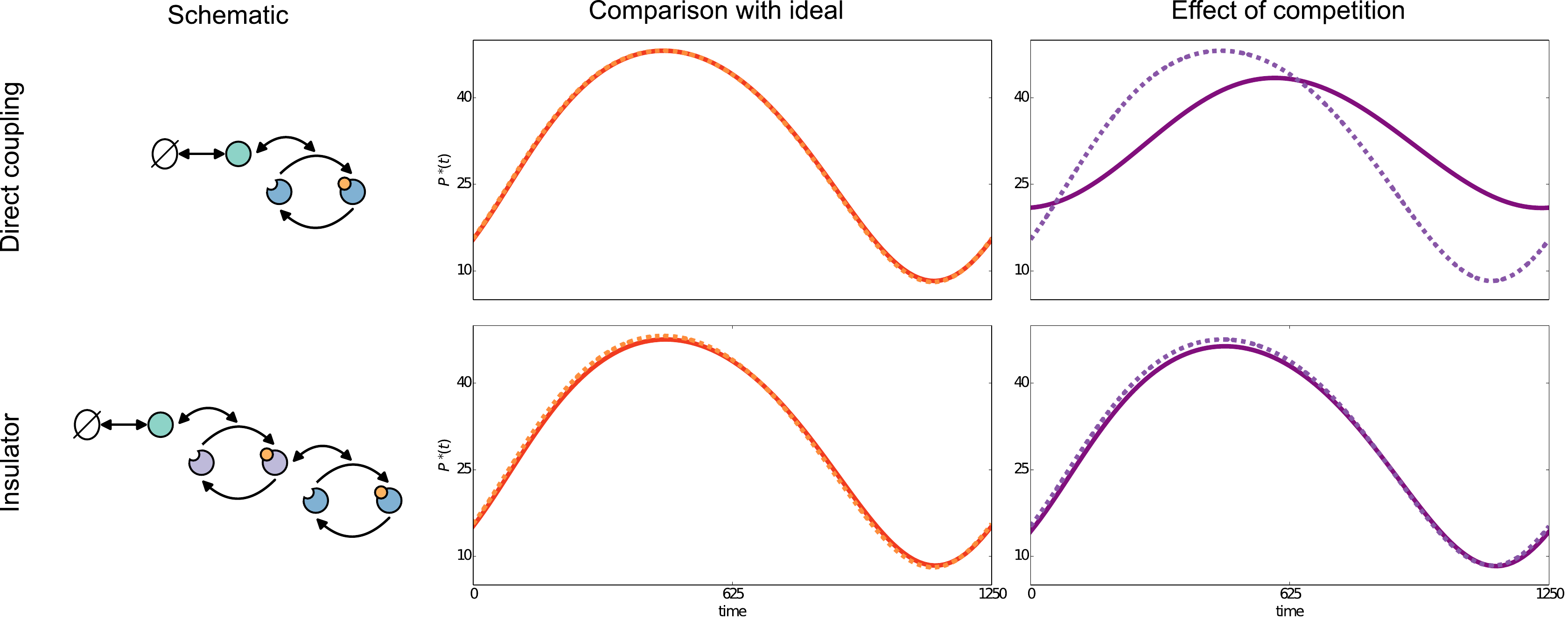}
\caption{{\small 
Retroactivity effects lead to signal distortion and attenuation of output signals when additional targets are added. 
Comparison of retroactivity effects on an enzymatic system with a direct coupling (DC) architecture (\textit{top}) and one with an insulator, represented by a phosphorylation/dephosphorylation cycle (\textit{bottom}).
Left column, a cartoon schematic of the enzymatic system. In the DC system \eqref{basic reactions}, the input interacts directly with the target. With an insulator \eqref{PD reactions}, the input drives phosphorylation of an intermediate signaling molecule, whose phosphorylated form then interacts with the target.
Middle column, illustration of {\dn}. The ``ideal'' output signal (\textit{dashed}), see \eqref{ideal ODEs}, with retroactivity effects neglected, is plotted against the output for each system with nonlinear dynamics (\textit{solid}), given by \eqref{basic ODEs} for the DC system and \eqref{PD ODEs} for the insulator.
Right column, illustration of {\ce}. The output signal in a system with a single target (\textit{dashed}) is compared with the output signal when multiple targets are present (\textit{solid}).
Plots of the output signals in each system are shown in the steady state, over a single period of $k(t)$. This plot was made using the parameters $k(t)=0.01\left(1+\sin{\left(0.005\,t\right)}\right)$, $\delta=\alpha_1=0.01$, $k_1=\alpha_2=10$, and $P_{\rm tot}=100$, with $X_{\rm tot}=Y_{\rm tot}=100$, $\beta_1=\gamma_1=0.01$, and $k_2=k_3=\beta_2=\gamma_2=10$ for the insulator. Parameters specifying the interaction with the new target P$^\prime$ in the perturbed system are $k^\prime=\alpha_2^\prime=10$, $\delta^\prime=\alpha_1^\prime=0.01$, and $P^\prime_{\rm tot}=10^4$.}
\label{fig:output}}
\end{center}
\end{figure}

For the enzymatic model considered here, both the DC system and the system with an insulator typically exhibit low distortion (Fig.~\ref{fig:output}). However, the use of an insulator results in a substantial reduction in competition effect. Indeed, even the inclusion of a parallel target P$^\prime$ with $P^\prime_{\rm tot}=10^4$, two orders of magnitude larger than the standard value $P_{\rm tot}=10^2$ used here, leads to only minor perturbations of the ouptut for the system including an insulator. In contrast, the DC system is strongly affected. 

Energy use is critical for optimal performance of the insulator. Metabolic processes ensure that the phosphorylation-dephosphorylation cycle is driven out of equilibrium, therefore consuming energy, by maintaining the ratio of concentrations of phosphate donors and acceptors, such as ATP and ADP, far from their equilibrium values. We measure the energy use of the insulator by the rate of ATP consumption in the phosphorylation-dephosphorylation cycle, which is just given by the average current through the phosphorylation step of the cycle
\begin{equation} \label{power}
J = \langle k_2 C_2 \rangle \, .
\end{equation}
Unlike the previous case considered in \citep{biophysj}, here the DC system, without the insulator, also consumes energy during the phosphorylation step. In the same way as for the insulator, we quantify this rate of ATP consumption as
\begin{equation} \label{power2}
J^\prime = \langle k_1 C_1 \rangle \, .
\end{equation}
In the following analysis we show, as in \citep{biophysj}, that better performance of the insulator requires more energy consumption.

\section{Results} \label{results}

\begin{figure}
\begin{center}
\hspace*{-10pt}
\includegraphics*[width=6.5cm]{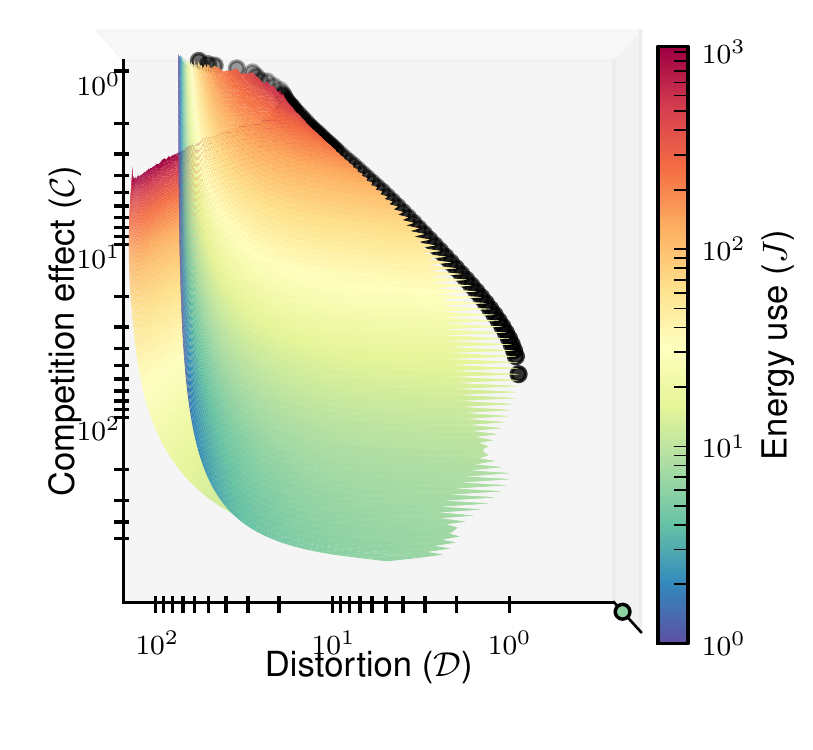}
\hspace*{12pt}
\includegraphics*[width=6.5cm]{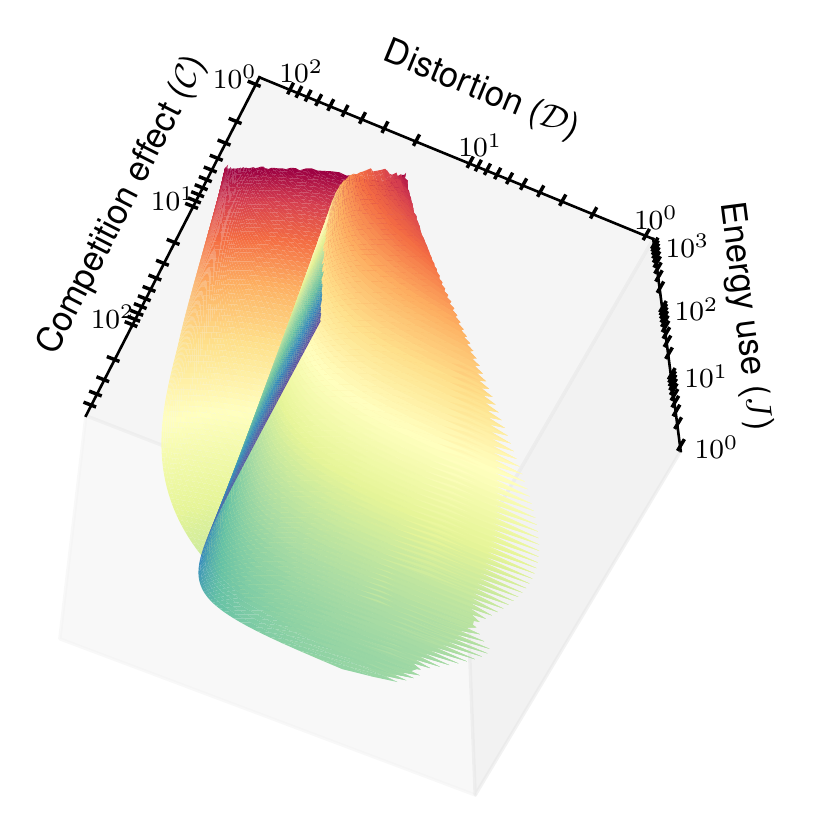}
\caption{\small{
Performance of the insulator measured by the {\ce} $\Comp$ and {\dn} $\Dist$ of the output in the system with an insulator \eqref{PD ODEs}, tested over a range of $X_{\rm tot}$ and $Y_{\rm tot}$ varied independently from $10$ to $10000$ in logarithmic steps. For simplicity $\Comp$ and $\Dist$ are rescaled such that the smallest (best) values are equal to one. Points are shaded according to the logarithm of the rate of the total rate of energy consumption $J_{\rm tot}=J+J^\prime$.
In the left plot, only $\Comp$ and $\Dist$ are plotted. Pareto efficient parameter points are marked by black dots. 
In the right plot, we show a three-dimensional view of the performance surface, also plotting the total rate of energy consumption. 
Generically, rates of energy consumption increase as one approaches the Pareto front; obtaining small values of the {\ce} is particularly costly. For comparison, $\Comp$ and $\Dist$ for the direct coupling system is marked by a dot in the left plot. The dot's shading reflects the rate of energy consumption $J^\prime$ for the DC system alone. See Section~\ref{results} for details. This plot was made using the same parameters as in Fig~\ref{fig:output}.}
\label{fig:Pareto}}
\end{center}
\end{figure}

We have tested the performance of the insulating PD cycle over an extensive range of parameters to explore the relationship between insulation, as measured by $\Comp$ and $\Dist$, and energy consumption rates. In Fig.~\ref{fig:Pareto} we show a plot of $\Comp$ and $\Dist$ for systems with a range of $X_{\rm tot}$ and $Y_{\rm tot}$ values, obtained by numerical integration of the differential equations \eqref{PD ODEs}. As a comparison we also show the values of $\Comp$ and $\Dist$ obtained for the DC system, as well as its rate of energy consumption $J^\prime$. As mentioned in Section~\ref{insulator}, the DC system in fact obtains a lower value for $\Dist$, and a higher value for $\Comp$, than observed for any system with an insulator. Surprisingly, for some values of the parameters the system with an insulator consumes less energy in total than the DC system, despite the introduction of an additional PD cycle.

We would like to associate the overall quality of the insulator performance with its energy use. However, minimization of competition effect and distortion is a multiobjective optimization problem, and no choice of parameters achieves the lowest value for $\Comp$ and $\Dist$ simultaneously. Thus we adopt a Pareto point of view for assessing the insulator performance, searching for points in parameter space where any improvement in $\Comp$ necessitates a sacrifice of $\Dist$, and vice versa. Such points in parameter space are referred to as Pareto optimal or Pareto efficient points. In the left plot of Fig.~\ref{fig:Pareto} the Pareto optimal choices of parameters on the tested parameter space are indicated by black points.

\begin{figure}
\begin{center}
\includegraphics*[width=9cm]{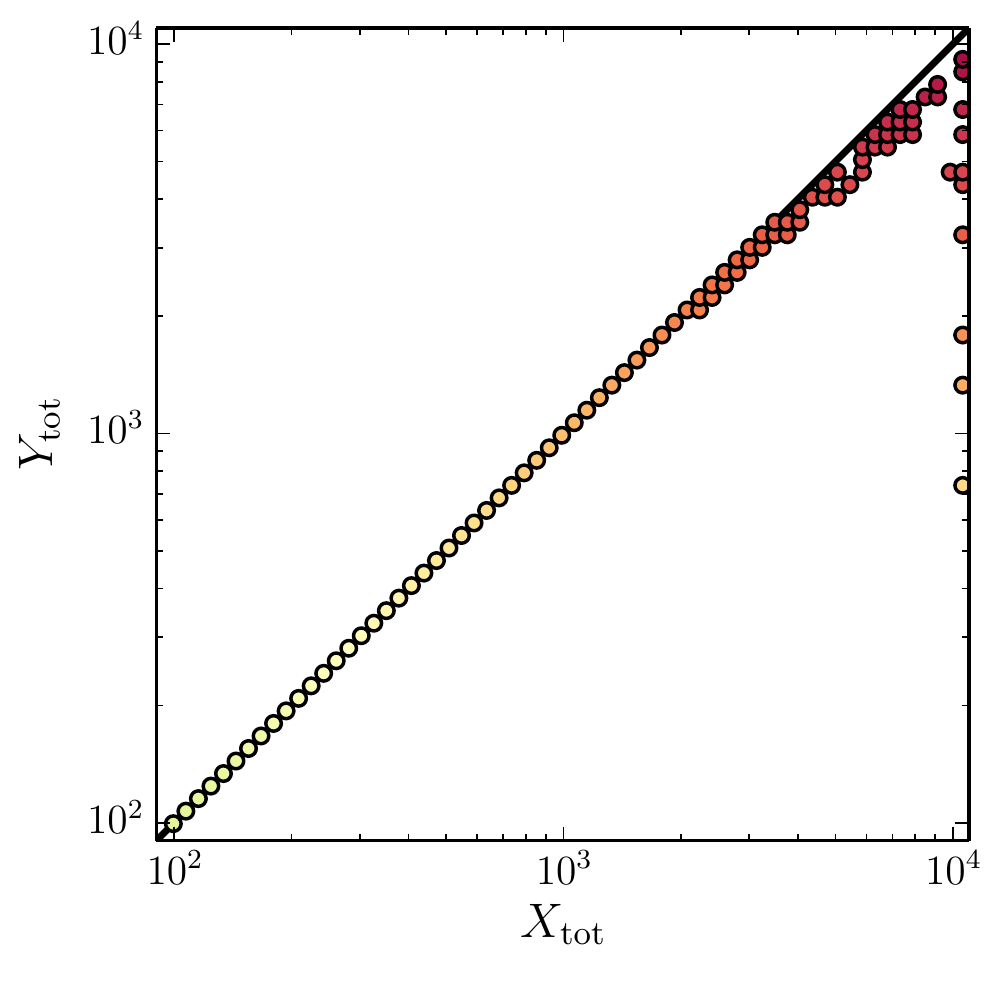}
\caption{
Scatter plot of the Pareto optimal sets of parameters $X_{\rm tot}$ and $Y_{\rm tot}$ corresponding to those in Fig.~\ref{fig:Pareto}. Pareto optimal points strike a balance between the total concentrations of X and Y (for reference the line $X_{\rm tot}=Y_{\rm tot}$ is shown in the background). Each point is shaded according to the rate of energy consumption for that choice of parameters. Increases in either $X_{\rm tot}$ or $Y_{\rm tot}$ result in increased energy expenditure. Due to the limited range of parameters which could be tested, some Pareto optima lie along the boundaries of the parameter space (see the ``elbow'' in the scatter points at the top right of the plot).
\label{fig:optima}}
\end{center}
\end{figure}

The shape of the performance space, parameterized by $\Comp$ and $\Dist$, is similar to that observed in \citep{biophysj}, as is the set of Pareto optimal parameters (see Fig.~\ref{fig:optima}). We also find that the rate of energy consumption increases as one approaches the line of Pareto optimal points, called the Pareto front, where performance of the insulator is optimal. Systems with parameters chosen on or near the Pareto front have some of the highest rates of energy expenditure observed, while systems which perform poorly tend to consume less energy.

\subsection*{Geometry of the optimal parameter space}

As described above, $\Comp$ and $\Dist$ quantify two competing objectives: robustness of the output to changes in downstream targets (measured by $\Comp$), and faithful signal transmission (measured by $\Dist$). Real insulators should optimize some combination of these quantities. As in \citep{biophysj} we found that insulators with minimal distortion have low values of $X_{\rm tot}$ and $Y_{\rm tot}$, and those with minimal competition effect have high $X_{\rm tot}$ and $Y_{\rm tot}$. Pareto optimal paremeter choices interpolate between these two extremes. 

These findings are similar to those given in a much broader context in~\citep{shoval2012}, where Shoval et al.~examined a simple model of biological systems that must satisfy multiple objectives. There, they described the phenotype of a biological system by a vector of traits (e.g.~body size and wing shape in bats). They observed that the values of certain traits of naturally-occurring phenotypes often lie on a low-dimensional subspace of the full trait space. This potentially surprising observation can be naturally explained within the framework of multi-objective optimization, where the phenotypes that satisfy the best tradeoffs between multiple objectives are weighted averages of certain ``archetypes'', which give the best performance on single tasks. 

\begin{figure}
\begin{center}
\includegraphics*[width=9cm]{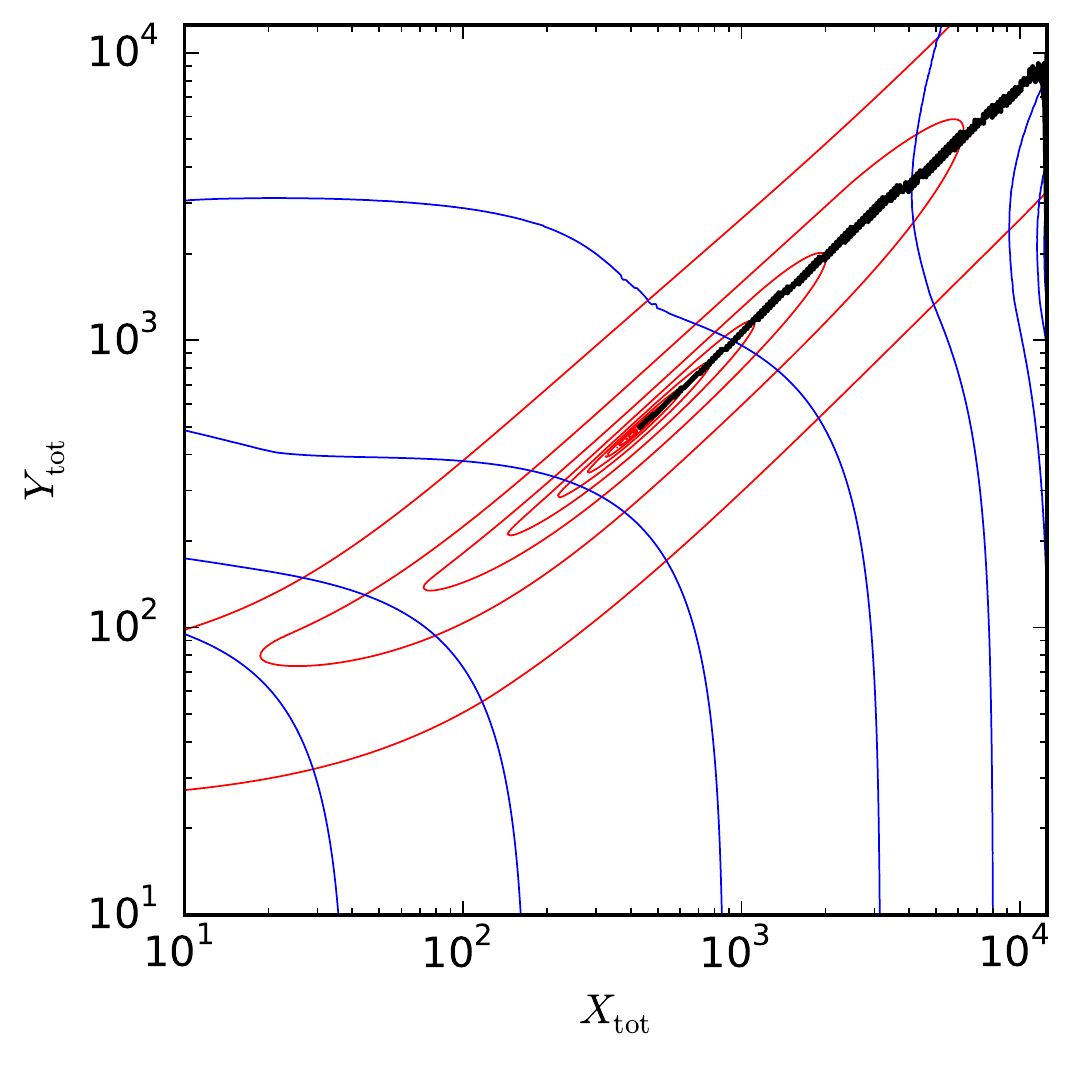}
\caption{
Level sets of $\Comp$ (red) and $\Dist$ (blue). Pareto optimal values of $X_{\rm tot}$ and $Y_{\rm tot}$ are marked by black dots.
\label{fig:contour}}
\end{center}
\end{figure}

It was shown that such results are generically obtained in the case that each objective is optimized by a single phenotype, and that performance decreases for each objective with increasing distance from the optimal phenotype \citep{shoval2012}. In the simple case that level sets of the objective functions are elliptical, the Pareto front is close to linear for two objectives. Here, we see that although the level sets of $\Comp$ and $\Dist$ are not simple ellipses (Fig.~\ref{fig:contour}) the Pareto front forms a line in the parameter space. Interestingly, in our case the Pareto optimal values also form a line in the space of $\Comp$ and $\Dist$ (see Fig.~\ref{fig:Pareto}). 

\bibliographystyle{unsrt}
\bibliography{refs}

\end{document}